**Simulation of Growth of Graded Bandgap Solid Solutions of $GaAs_xP_{1-x}$ at Liquid Phase Electroepitaxy**

V.V. Tsybulenko, Ye.A. Baganov, V.A. Krasnov, S.V. Shutov
Kherson National Technical University,
Department of Energy and Electrical Engineering
24, Beryslavs'ke highway, Kherson, Ukraine, 73008,
Phone: +38 (0552) 516468, e-mail: phisel@mail.ru

## *1. Introduction*

Solid solutions of $A^3B^5$ with the spatial dependence of the bandgap and, consequently, with built-in electrical fields are widely used in optoelectronics. Particularly, for effective photovoltaic converters it is used the graded bandgap of the $GaAs_{1-x}P_x$ solid solutions with the bandgap that increases towards the surface of the emitter layer [1].

Obtaining of such structures by the methods of epitaxy from the liquid phase allows to improve considerably in operating characteristics due to higher crystal quality of epitaxial layers [2].

The conventional methods of the liquid phase epitaxy (LPE), that use forced cooling or isothermal growth from the supersaturated solution-melt do not provide the possibility to form the epitaxial layers with the bandgap that increases from the interface of the substrate/emitter layer towards the surface of the emitter layer [3].

Using of the liquid phase electroepitaxy (LPEE) for the growth of the graded bandgap epitaxial layers gives the following advantages. Firstly, LPEE gives the possibility to use the isothermal conditions of the growth, which are more suitable for the technological process. Secondly, application of the electric current allows to decrease considerably the response time of the epitaxial process and to provide the possibility of additional control at the epitaxy.

In the present article the possibility of control of the composition of the solid solution by the example of the $GaAs_{1-x}P_x$ solid solution at the LPEE from the Ga-As-P solution-melt on the GaAs substrate is theoretically considered.

## 2. *Mathematical model*

The simulated growing cell is given in the Fig. 1. Initially the saturated As-solution in the Ga-melt is brought to a contact with the GaAs substrate (namely working substrate) and the GaP substrate (namely source substrate) that is the source of the phosphorous for the solution-melt. The substrates are situated parallel to each other at the certain distance.

At the modeling there are following assumptions have been used:

- Growth process is isothermal;
- Mechanical stresses are not considered;
- It is considered diffusion processes of mass transfer are present in the solution-melt only;
- It is considered all growth and dissolution processes are diffusion limited;
- Processes of transfer of the components in the solution-melt are independent;
- Interfaces of solid phases – solution-melt are considered being unmovable hence the Stefan problem does not take place.

According to the assumption about the diffusion limitation of the crystallization process, the concentrations of the components of the solid and liquid phases near the interfaces have to be in equilibrium. For the ternary system of the $GaAs_{1-x}P_x$ solid solution the equations that describe solidus and liquidus lines at the regular solution approach are as follows [4]:

$$x = \left[ \frac{4x_{0Ga}^l x_{0P}^l \gamma_{Ga}^l \gamma_P^l}{\gamma_{Ga}^{sl} \gamma_P^{sl}} \cdot \exp\left( \frac{\Delta S_{GaP}^F \left( T_{GaP}^F - T \right)}{RT} \right) \right] \Big/ \gamma_{GaP}^s \tag{1}$$

$$(1-x) = \left[ \frac{4x_{0Ga}^l x_{0As}^l \gamma_{Ga}^l \gamma_{As}^l}{\gamma_{Ga}^{sl} \gamma_{As}^{sl}} \cdot \exp\left( \frac{\Delta S_{GaAs}^F \left( T_{GaAs}^F - T \right)}{RT} \right) \right] \Big/ \gamma_{GaAs}^s \tag{2}$$

where $x = x_{GaP}^s$ is the mole fraction of GaP in the $GaP_xAs_{1-x}$ solid solution; $x_{0Ga}^l, x_{0As}^l, x_{0P}^l$ are the mole fractions of Ga, As, and P respectively in the liquid phase, that correspond to the equilibrium with the solid phase at $z = 0$; $\gamma_{Ga}^l, \gamma_{As}^l, \gamma_P^l$ are the activity coefficients of Ga, As, and P respectively in the liquid phase; $\gamma_{Ga}^{sl}, \gamma_{As}^{sl}, \gamma_P^{sl}$ are the activity coefficients of Ga, As, and P respectively in the liquid phase of stoichiometric composition; $\gamma^s_{GaAs}, \gamma^s_{GaP}$ are the activity coefficients of GaAs and GaP respectively in the solid phase; $\Delta S^F_{GaAs}, \Delta S^F_{GaP}$ are the specific mole entropies of the melting of GaAs and GaP respectively; $T^F_{GaAs}, T^F_{GaP}$ are the temperatures of melting of GaAs and GaP respectively; $T$ is the temperature of the epitaxy; $R$ is the absolute gas constant.

Activity coefficients can be calculated as follows [4]:

$$\gamma^l_{Ga} = \exp\left(\frac{\alpha^l_{GaP}\left(x^l_{0P}\right)^2 + \alpha^l_{GaAs}\left(x^l_{0As}\right)^2 + \left(\alpha^l_{GaP} + \alpha^l_{GaAs} - \alpha^l_{AsP}\right)x^l_{0As}x^l_{0P}}{RT}\right) \tag{3}$$

$$\gamma^l_{As} = \exp\left(\frac{\alpha^l_{AsP}\left(x^l_{0P}\right)^2 + \alpha^l_{GaAs}\left(x^l_{0Ga}\right)^2 + \left(\alpha^l_{AsP} + \alpha^l_{GaAs} - \alpha^l_{GaP}\right)x^l_{0Ga}x^l_{0P}}{RT}\right) \tag{4}$$

$$\gamma^l_{P} = \exp\left(\frac{\alpha^l_{GaP}\left(x^l_{0Ga}\right)^2 + \alpha^l_{AsP}\left(x^l_{0As}\right)^2 + \left(\alpha^l_{GaP} + \alpha^l_{AsP} - \alpha^l_{GaAs}\right)x^l_{0Ga}x^l_{0As}}{RT}\right) \tag{5}$$

$$\gamma^{sl}_{Ga} = \gamma^{sl}_{As} = \exp\left(\frac{0.25\alpha^l_{GaAs}}{RT}\right),\ \gamma^{sl}_{Ga} = \gamma^{sl}_{P} = \exp\left(\frac{0.25\alpha^l_{GaP}}{RT}\right) \tag{6}$$

$$\gamma^s_{GaAs} = \exp\left(\frac{\alpha^s_{GaAs-GaP}x^2}{RT}\right),\ \gamma^s_{GaP} = \exp\left(\frac{\alpha^s_{GaAs-GaP}(1-x)^2}{RT}\right) \tag{7}$$

where $\alpha^l_{GaAs}, \alpha^l_{GaP}, \alpha^l_{AsP}$ are the parameters of atoms interaction in the liquid phase for Ga-As, Ga-P, and As-P respectively; $\alpha^s_{GaAs-GaP}$ is the parameter of atoms interaction in the solid phase for GaAs- GaP.

Diffusion processes of the As and P components soluted in the Ga-melt are described as follows [5, 6, 7]:

$$\frac{\partial x^l_{As}(z,t)}{\partial t} = -\mu_{As}E_z\frac{\partial x^l_{As}(z,t)}{\partial z} + D^l_{As}\frac{\partial^2 x^l_{As}(z,t)}{\partial z^2} \tag{8}$$

$$\frac{\partial x_P^l(z,t)}{\partial t} = -\mu_P E_z \frac{\partial x_P^l(z,t)}{\partial z} + D_P^l \frac{\partial^2 x_P^l(z,t)}{\partial z^2} \qquad (9)$$

where $\mu_{As}, \mu_P, D_{As}^l, D_P^l$ are the electrical mobility and diffusivity of ions of As and P in the Ga-melt respectively; $x_{Ga}^l(z,t), x_{As}^l(z,t), x_P^l(z,t)$ are the mole fractions of the components of Ga, As, and P respectively, that depends on time *t* and coordinate *z*.

Electric field can be found from the formula

$$E_z = J \cdot \rho_{Ga}^l \qquad (10)$$

where $\rho_{Ga}^l$ is the specific resistance of the solution-melt; *J* is the current density.

Due to the diffusion limited approach of the processes of the growth and the dissolution the boundary conditions are as follows:

$$\begin{array}{l} x_P^l(0,t) = x_{0P}^l;\ x_{As}^l(0,t) = x_{0As}^l \\ x_P^l(l,t) = x_{lP}^l;\ x_{As}^l(l,t) = x_{lAs}^l \end{array}, \qquad (11)$$

where $x_{lGa}^l, x_{lAs}^l, x_{lP}^l$ are the mole fractions of Ga, As, and P respectively in the liquid phase, that correspond to the equilibrium with the solid phase at $z = l$.

Initial conditions:

$$x_P^l(z,0) = x_{0P}^l;\ x_{As}^l(z,0) = x_{0As}^l \qquad (12)$$

For the solving of the equations of mass transfer (8, 9) with the initial and the boundary conditions (11, 12) the method of finite differences was used. The explicit scheme of approximation of equations (8, 9) was used [8]. Boundary conditions (11) were considered as follows

At the interface $z = 0$ due to the crystallization the composition of the solid phase changes in time, that leads to implicit dependence of $x_{0Ga}^l, x_{0As}^l$, and $x_{0P}^l$ on time.

The composition of the layer crystallizing during a time step can be described as follows:

$$x = \frac{1}{\dfrac{x^{*l}_{As} - x_{0As}^l}{x^{*l}_{P} - x_{0P}^l} + 1} \qquad (13)$$

where $x^{*l}_{As}, x^{*l}_{P}$ are the mole fractions of As and P in the liquid phase respectively, that have been appeared at current time step due to the mass transfer in the liquid phase at $z = 0$ before crystallization.

Solving the system of the three equations (1, 2, 13) and the normalization requirement $x_{0Ga}^{l} = 1 - \left(x_{0As}^{l} + x_{0P}^{l}\right)$ it can be obtained the equilibrium composition of the liquid and solid phases, i.e. $x, x_{0Ga}^{l}, x_{0As}^{l}, x_{0P}^{l}$

The thickness of layer crystallizing during a time step can be found as follows:

$$h^{g} = \frac{h}{\omega^{s}}\left(\frac{\left(x^{*l}_{As} - x_{0As}^{l}\right) + \left(x^{*l}_{P} - x_{0P}^{l}\right)}{\left(x_{0Ga}^{l}\frac{M_{Ga}}{\rho_{Ga}} + x_{0As}^{l}\frac{M_{As}}{\rho_{As}} + x_{0P}^{l}\frac{M_{P}}{\rho_{P}}\right)}\right). \quad (14)$$

where $\omega^{s} = \dfrac{8}{N_{A}\left[x a_{GaP} + (1-x) a_{GaAs}\right]^{3}}$ (15)

is the mole density [6], $N_A$ is the Avogadro constant; $a_{GaP}, a_{GaAs}$ is the lattice constants of GaP and GaAs respectively; $M_{Ga}, M_{As}, M_{P}, \rho_{Ga}, \rho_{As}, \rho_{P}$ are the molar masses and densities of Ga, As, and P respectively; $h$ is the spatial step for the finite differences scheme.

The discretization of the boundary conditions (11) at the interface substrate GaP/solution-melt ($z = l$) be considered as follows. By the law of conservation of mass:

$$\left(x^{**l}_{Ga} + x^{**l}_{As} + x^{**l}_{P}\right) + \alpha \cdot GaP = x_{lGa}^{l} + x_{lAs}^{l} + x_{lP}^{l} \quad (16)$$

where $x^{**l}_{Ga}, x^{**l}_{As}, x^{**l}_{P}$ are the mole fractions of Ga, As, and P respectively in the liquid phase, that appeared at current time step due to mass transfer in the liquid phase at $z = l$ before dissolution.

The mole fraction of Ga can be found from the normalization requirement:

$$x_{lGa}^{l} = 1 - \left(x_{lAs}^{l} + x_{lP}^{l}\right) \quad (17)$$

$$x^{**l}_{Ga} = 1 - \left(x^{**l}_{As} + x^{**l}_{P}\right) \quad (18)$$

Form the equations (16)-(18) it can be obtained:

$$\begin{cases} \dfrac{1-\left(x^{**l}{}_{As}+x^{**l}{}_{P}\right)+\dfrac{1}{2}\alpha}{1+\alpha}=1-\left(x_{lAs}^{l}+x_{lP}^{l}\right) \\ \dfrac{x^{**l}{}_{P}+\dfrac{1}{2}\alpha}{1+\alpha}=x_{lP}^{l} \end{cases} \tag{19}$$

As a solution of a system (19) on can obtain:

$$x_{lP}^{l}=\frac{0.5\left(2\cdot x^{**l}{}_{P}\cdot x_{lAs}^{l}+x^{**l}{}_{As}-x_{lAs}^{l}\right)}{x^{**l}{}_{As}} \tag{20}$$

Solving the system includes two equations of phase equilibrium (1, 2), normalization requirement (18) and equation (20) at every time step of the calculations gives the equilibrium compositions of the liquid and solid phases at interface of GaP substrate/solution-melt, i.e. $x_{lGa}^{l}, x_{lAs}^{l}, x_{lP}^{l}$

## *3. Results and discussion*

The simulation was carried out in the range of temperatures of 600-900$^{o}$C, within which all range of $GaAs_{1-x}P_{x}$ solid solutions is present [9], in the range of distances between substrates (thickness of the growth space) of 0.5-1 mm, and in the range of current densities of 1-9 A/cm$^2$.

In the Table 1 the values of parameters, been used in calculations are given.

The figures 2a – 2c evident as follows:

- Under steady-state conditions a value of the current density has influence on the growth rate of the layer of solid solution only and practically has no influence on (fig. 2a);
- Under steady-state conditions the gradient of composition of the solid solution depends on the temperature of growth (fig. 2b) and on the distance between the substrates (fig. 2c). Decreasing of the temperature of growth or the distance between the substrates leads to increasing of the composition gradient.

As appears from the figure 3, under steady-state conditions changing of such parameters as the temperature of growth and/or the thickness of the growth space gives the possibility to control the gradient of the composition of the $GaAs_{1-x}P_x$ solid solution in a wide range from $0.5\cdot10^{-4}$ mole. frac./nm to $2.0\cdot10^{-3}$ mole. frac./nm.

Let's consider an influence of change of the current density on processes in the system. In the figures 4a and 4b the evolutions of concentrations of P and As atoms in Ga-melt at current switches off are given.

Before the current switching off considerable difference in profiles of concentrations of P and As is caused by different directions of electric and diffusion parts of the As and P atoms flows. For P both electric and diffusion parts of the atoms flow have the same opposite direction to the 0z axe. But for As the diffusion part of the atoms flow has the 0z axe and the electric one has the opposite direction. That is why the profile of distribution of As concentration depends on the value of the current density to a greater extent than distribution of P concentration. Changes in current density lead to a considerable redistribution of As concentration in the liquid phase and, as a result, at the crystallization front. It provides shift of equilibrium conditions in the system and moves the figurative work point in the phase diagram.

Switching off the current leads to the decreasing of As concentration at the crystallization front due to the diffusion. It results in increasing of mole fraction of P in the solid phase.

Thus just ***changing*** of magnitude of current density can be used for the control of the gradient of the composition of the solid solution during growth (see fig. 5.).

## *Conclusions*

1. It was shown that use of electroepitaxy from the liquid phase in the isothermal conditions gives the principal possibility of obtaining of graded bandgap $GaAs_{1-x}P_x$ structures with increasing of content of the phosphorous toward the surface of the layer.
2. It was shown that use of unsteady electric field provides the possibility of effective control of composition of the ternary solid solution.

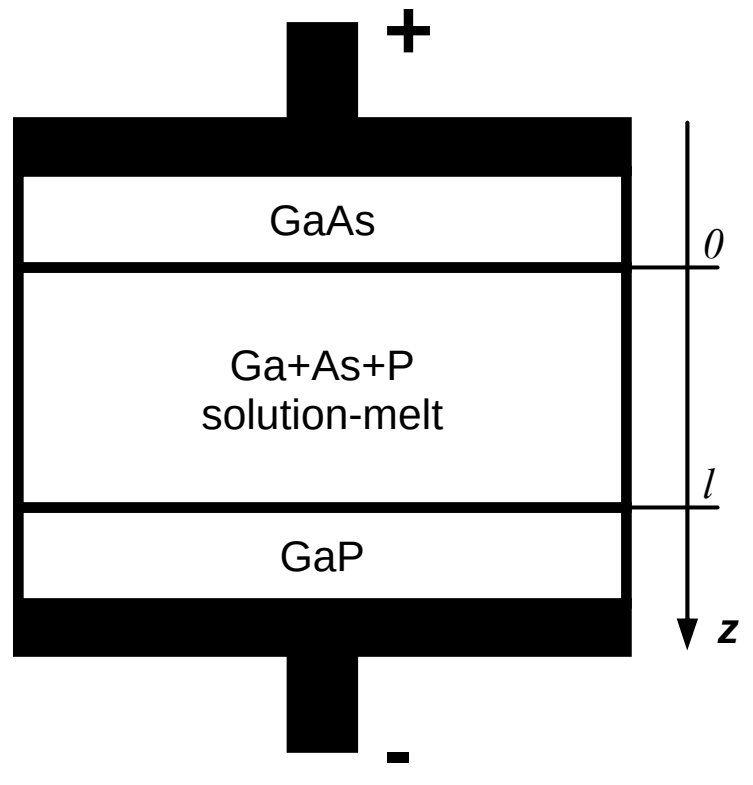

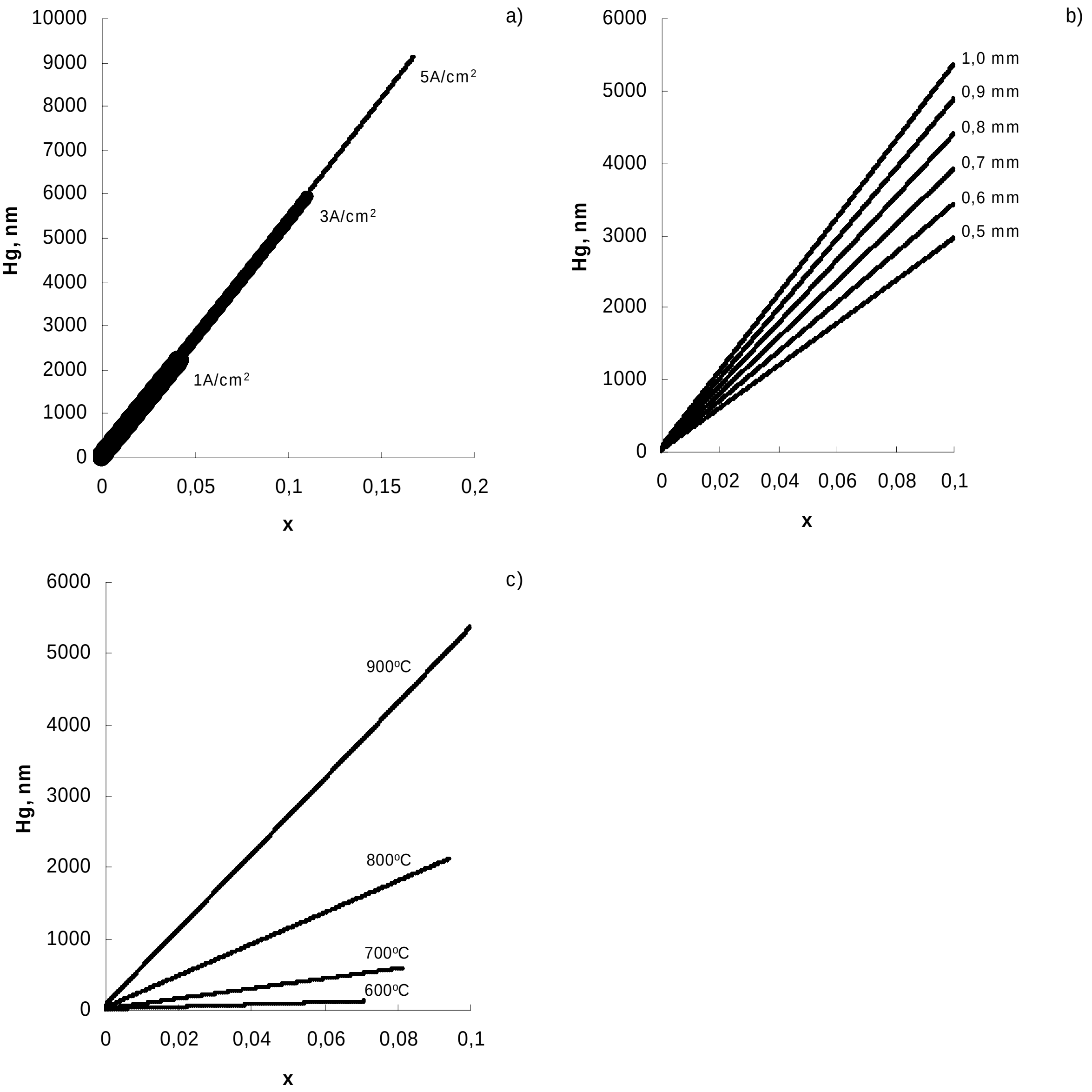

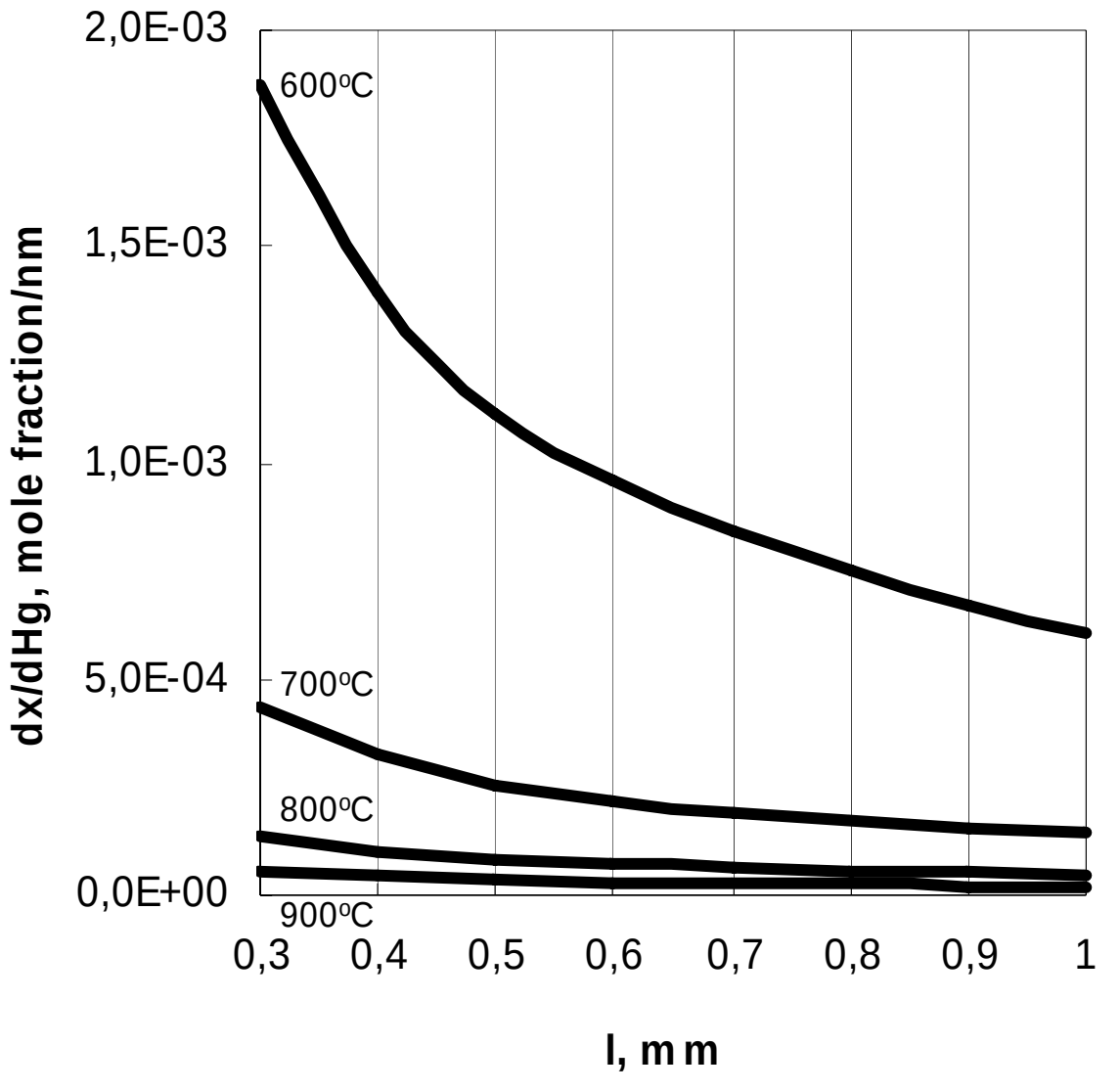

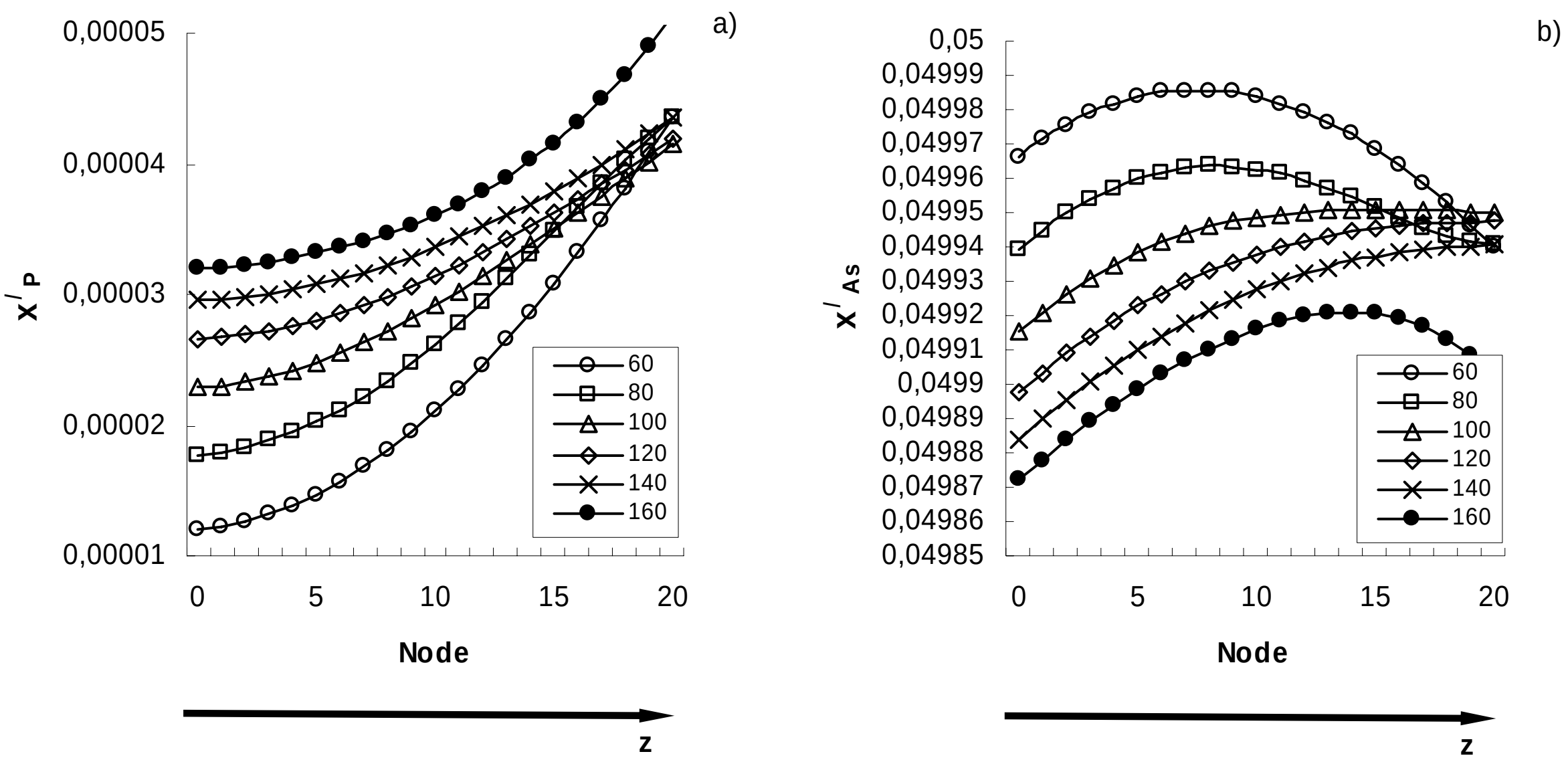

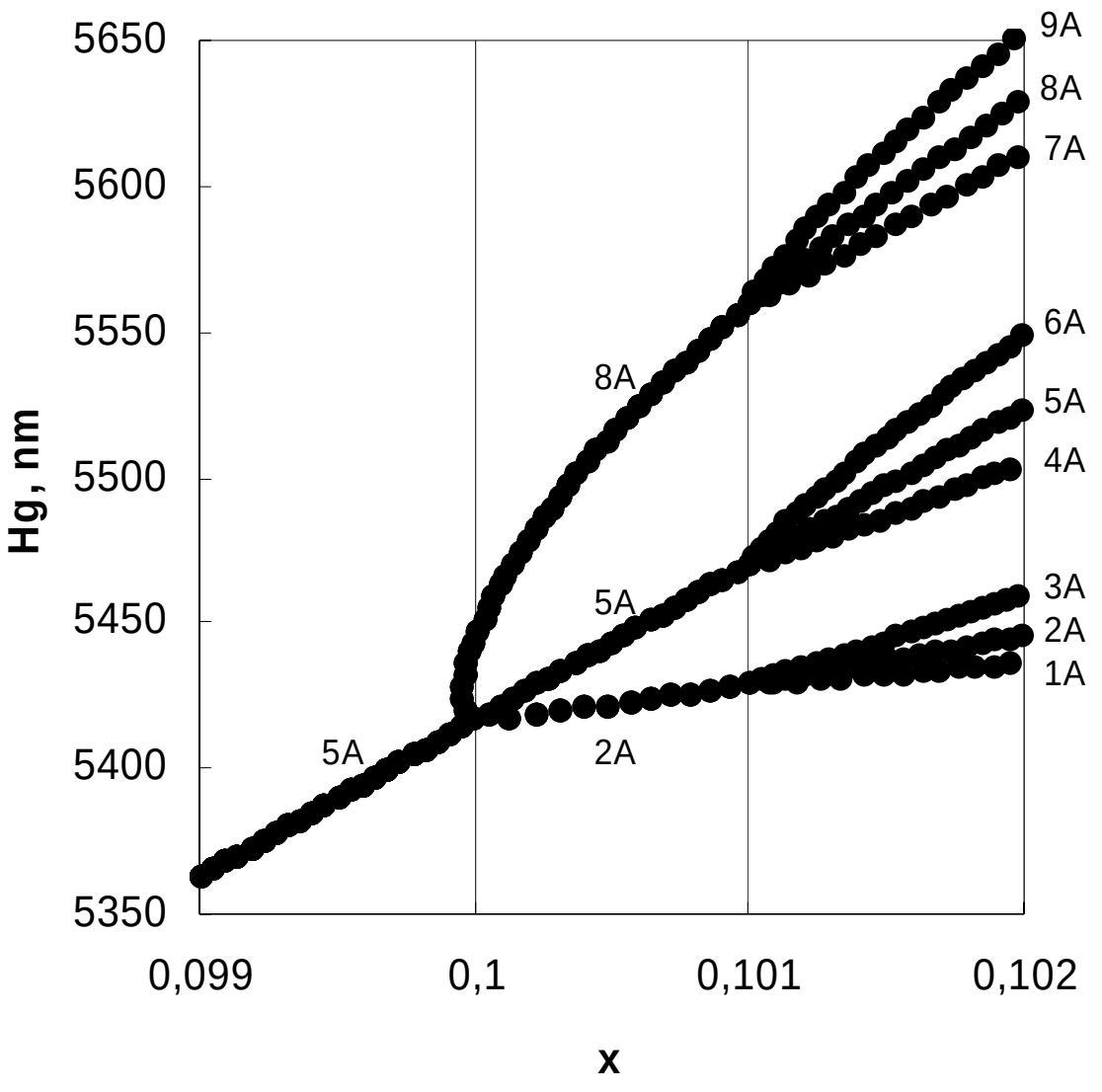

***Figure captions***

Figure 1. The growing cell for LPEE.

Figure 2. Dependence of the thickness of the layer of solid solution $GaAs_{1-x}P_x$ being deposited on its composition:

*a)* at different densities of current, for the same time of an hour (the temperature of the growth $T = 900^oC$, the thickness of the growth space $l$ =1 mm);

*b)* at different thicknesses of the growth space (the temperature of the growth $T = 900^oC$, the density of current $J$ =3 A/cm$^2$);

*c)* at different temperatures of the growth (the thickness of the growth space $l$ = 1 mm, the density of current $J$ =3 A/cm$^2$).

Figure 3. Dependence of the gradient of the composition of the $GaAs_{1-x}P_x$ solid solution on parameters of growth process: the temperature of growth and the thickness of the growth space. The density of current $J$ =3 A/cm$^2$.

Figure 4. The node distribution of the mole fraction of the P (a) and As (b) at different time. The temperature of growth is 900$^o$C, the thickness of the growth space is 1 mm. The current density: till 60 s it was 1 A/cm$^2$, in the range of 80-140 s it was switched off (0 A/cm$^2$), and at 140 s it was switched on with the previous magnitude of 1 A/cm$^2$.

Figure 5. Dependence of the thickness of the growing layer of the $GaAs_{1-x}P_x$ solid solution on its composition at current density, that changes intended during the growth and represents unsteady electric field (the temperature of growth is 900$^o$C and the thickness of the growth space is 1 mm).

Table 1

| *Designation* | *Description* | *Value* | *Source* |
|---|---|---|---|
| $\mu_{As}, \frac{cm^2}{V \cdot s}$ | the electrical mobility of the As ions in Ga | $16 \cdot 10^{-3}$ | [10, 7] |
| $\mu_{P}, \frac{cm^2}{V \cdot s}$ | the electrical mobility of the P ions in Ga | $36.1 \cdot 10^{-3}$ | [11] |
| $D^l_{As}, \frac{cm^2}{s}$ | diffusivity of the As atoms in Ga. | $4.229 \cdot 10^{-5}$ | [12] |
| $D^l_{P}, \frac{cm^2}{s}$ | diffusivity of the P atoms in Ga. | $4.798 \cdot 10^{-5}$ | [11,12] |
| $\Delta S^F_{GaAs}, \frac{cal}{mole \cdot {}^o C}$ | the specific mole entropy of the melting of GaAs | 16.64 | [3,4,13] |
| $\Delta S^F_{GaP}, \frac{cal}{mole \cdot {}^o C}$ | the specific mole entropy of the melting of GaP. | 15.0 | [3,4,13] |
| $T^F_{GaAs}, K$ | the temperature of melting of GaAs. | 1511 | [3,4,13] |
| $T^F_{GaP}, K$ | the temperature of melting of GaP. | 1743 | [3,4,13] |
| $\alpha^l_{GaAs}, \frac{cal}{mole}$ | the parameter of atoms interaction in the liquid phase for Ga-As. | $-3.7 \cdot T$ | [3,4,5,14] |
| $\alpha^l_{GaP}, \frac{cal}{mole}$ | the parameter of atoms interaction in the liquid phase for Ga-P. | $7900 - 7 \cdot T$ | [3,4,13] |
| $\alpha^l_{AsP}, \frac{cal}{mole}$ | the parameter of atoms interaction in the liquid phase for As-P. | 2000 | [3,4,13] |
| $\alpha^s_{GaAs-GaP}, \frac{cal}{mole}$ | the parameter of atoms interaction in the solid phase for GaAs- GaP. | 1000 | [3,4,13] |
| $a_{GaAs}, cm$ | the lattice constant of GaAs. | $0.5641 \cdot 10^{-7}$ | [4,14] |
| $a_{GaP}, cm$ | the lattice constant of GaP. | $0.5449 \cdot 10^{-7}$ | [4,14] |
| $\rho^l_{Ga}, Ohm \cdot cm$ | the specific resistance of Ga. | $13.6 \cdot 10^{-6}\left(1 + 3.9 \cdot 10^{-3} \cdot T\right)$ | [14] |
| $\rho_{Ga}, \frac{g}{cm^3}$ | the Ga density | 6.09 | [12] |
| $\rho_{As}, \frac{g}{cm^3}$ | the *As density* | 5.72 | [12] |
| $\rho_{P}, \frac{g}{cm^3}$ | the P density | 2.0 | [12] |

**Моделювання процесів росту шарів $GaAs_xP_{1-x}$ при рідиннофазній електроепітаксії**

В.В. Цибуленко, Є.О. Баганов, В.О. Краснов, С.В. Шутов

Анотація

Теоретично розглянуто можливість управління складом твердого розчину $GaAs_{1-x}P_x$ на підкладці GaAs при рідиннофазній електроепітаксії з розчину-розплаву Ga-As-P. За допомогою методів математичного моделювання було встановлено, що в стаціонарних умовах, задаючи такі параметри процесу як температура та/або товщина ростового зазору, можна отримати варизонні шари твердого розчину $GaAs_{1-x}P_x$ із збільшенням вмісту P до поверхні шару, що мають градієнт складу від $0.5 \cdot 10^{-4}$ мольн.частка/нм до $2.0 \cdot 10^{-3}$ мольн.частка/нм. Показано, що керування складом тернарного твердого розчину в процесі рідиннофазної електроепітаксії може бути здійснене при використанні нестаціонарного електричного поля.

**Моделирование процессов роста слоев $GaAs_xP_{1-x}$ при жидкофазной электроэпитаксии**

В.В. Цыбуленко, Е.А. Баганов, В.А. Краснов, С.В. Шутов

Аннотация

Теоретически рассмотрена возможность управления составом твердого раствора $GaAs_{1-x}P_x$ на подложке GaAs при жидкофазной электроэпитаксии из раствора-расплава Ga-As-P. При помощи методов математического моделирования было установлено, что в стационарных условиях, задавая такие параметры как температура и/или толщина ростового зазора, можно получить варизонные слои твердого раствора $GaAs_{1-x}P_x$ с увеличением содержания P к поверхности слоя. Которое имеют градиент состава от $0.5 \cdot 10^{-4}$ мольн.доля/нм до $2.0 \cdot 10^{-3}$ мольн.доля/нм. Показано, что управление составом тернарного твердого раствора в процессе жидкофазной эпитаксии может бать реалізовано при использовании нестаціонарного электрического поля.

**Simulation of Growth of Graded Bandgap Solid Solutions of $GaAs_xP_{1-x}$ at Liquid Phase Electroepitaxy**

V.V. Tsybulenko, Ye.A. Baganov, V.A. Krasnov, S.V. Shutov

Summary

The possibility of the composition control of the $GaAs_{1-x}P_x$ solid solution on the GaAs substrate at liquid phase electroepitaxy from the Ga-As-P solution-melt is theoretically considered. By the simulation it was determined, that under steady-state conditions specifying such parameters of the process as the temperature and/or the thickness of the growth space it is possible to obtain graded bandgap layers of the $GaAs_{1-x}P_x$ solid solution with increasing of the content of P towards the surface of the layer that possess the composition gradient from $0.5\cdot10^{-4}$ mole fraction/nm to $2.0\cdot10^{-3}$ mole fraction/nm. It was also shown that control of the composition of ternary solid solutions at liquid phase electroepitaxy can be realized by use of unsteady state electric field.